\documentclass[letter,useAMS,usenatbib]{mn2e}
\usepackage{amsmath,bm}
\usepackage{graphicx}
\usepackage{enumerate}
\usepackage{bm}
\usepackage{times}
\usepackage{natbib}
\usepackage{url}
\usepackage{bm}
\graphicspath{{./figs/}{./png/}}

\newcommand{\etat}{\eta_{\rm t}}

\newcommand{\etaT}{\eta_{\rm T}}
\newcommand{\diff}{\rm{cm}^2\!/s}

{}

\newcommand{\rsun}{R}
\newcommand{\thetaI}{\theta_{\rm i}}

\newcommand{\itover}[2]{\,\hspace{-.15mm}#1{\!\hspace{.15mm}#2}}

\newcommand{\ithat}[1]{\itover{\hat}{#1}}

\newcommand{\deriv}[3]{\frac{#3\hspace*{-.06em} {#1}}{#3\hspace*{.06em} {#2}}}

\newcommand{\parder}[2]{\deriv{#1}{#2}{\partial}}

\newcommand{\EQ}{\begin{equation}}
\newcommand{\EN}{\end{equation}}
\newcommand{\EQA}{\begin{eqnarray}}
\newcommand{\ENA}{\end{eqnarray}}

\newcommand{\Eq}[1]{Equation~(\ref{#1})}

\newcommand{\bra}[1]{\langle #1\rangle}

\newcommand{\mean}[1]{\overline #1}

{}
{}
{}

{}
{}
\newcommand{\meanEMF}{\overline{\mbox{\boldmath ${\cal E}$}}{}}{}
{}
{}
{}
{}
{}
{}
{}
{}
{}
{}
{}
{}
{}
{}
{}
{}
{}

{}
{}
{}

{}

{}
{}

\newcommand{\rr}{\mbox{\boldmath $r$} {}}

\newcommand{\UU}{\mbox{\boldmath $U$} {}}

\newcommand{\BB}{\mbox{\boldmath $B$} {}}

\newcommand{\AAA}{\mbox{\boldmath $A$} {}}

\newcommand{\ee}{\mbox{\boldmath $e$} {}}

\newcommand{\nab}{\mbox{\boldmath $\nabla$} {}}

\def\la{\mathrel{\mathchoice {\vcenter{\offinterlineskip\halign{\hfil
$\displaystyle##$\hfil\cr<\cr\sim\cr}}}
{\vcenter{\offinterlineskip\halign{\hfil$\textstyle##$\hfil\cr<\cr\sim\cr}}}
{\vcenter{\offinterlineskip\halign{\hfil$\scriptstyle##$\hfil\cr<\cr\sim\cr}}}
{\vcenter{\offinterlineskip\halign{\hfil$\scriptscriptstyle##$\hfil\cr<\cr\sim\cr}}}}}
\def\ga{\mathrel{\mathchoice {\vcenter{\offinterlineskip\halign{\hfil
$\displaystyle##$\hfil\cr>\cr\sim\cr}}}
{\vcenter{\offinterlineskip\halign{\hfil$\textstyle##$\hfil\cr>\cr\sim\cr}}}
{\vcenter{\offinterlineskip\halign{\hfil$\scriptstyle##$\hfil\cr>\cr\sim\cr}}}
{\vcenter{\offinterlineskip\halign{\hfil$\scriptscriptstyle##$\hfil\cr>\cr\sim\cr}}}}}
\def\etat{\eta_{\rm t}}

\def\etaT{\eta_{\rm T}}

\newcommand{\Mm}{\,{\rm Mm}}

\renewcommand{\Eq}[1]{Eq.~(\ref{#1})}

\title[Plasma flow vs.\ magnetic feature-tracking speeds]
{Plasma flow vs.\ magnetic feature-tracking speeds in the Sun}
\author[Guerrero et al.]{G. Guerrero$^{1}$\thanks{E-mail:
guerrero@nordita.org (GG)}, M. Rheinhardt$^{1}$,
A. Brandenburg$^{1,2}$, and M. Dikpati$^{3}$\\
$^{1}$NORDITA, AlbaNova University Center,
Roslagstullsbacken  23, SE-10691 Stockholm, Sweden \\
$^{2}$Department of Astronomy, AlbaNova University Center,
Stockholm University, SE-10691 Stockholm, Sweden\\
$^{3}$High Altitude Observatory, National Center for Atmospheric
  Research \footnote{The National Center or Atmospheric Research 
    is sponsored by the National Science Foundation. }, 3080 Center 
  Green, Boulder, Colorado 80301}
\begin{document}

\date{Accepted 1988 December 15. Received 1988 December 14; in original form 1988 October 11, $ $Revision: 1.122 $ $}

\pagerange{\pageref{firstpage}--\pageref{lastpage}} \pubyear{2002}

\maketitle

\label{firstpage}

\begin{abstract}
We simulate the magnetic feature tracking (MFT) speed using advective-diffusive
transport models in both one and two dimensions. By depositing
magnetic bipolar regions
at different latitudes at the Sun's surface and
following their evolution for a prescribed
meridional circulation 
and magnetic diffusivity profiles, we derive the MFT speed as a function of
latitude. We find that in a one dimensional surface-transport model
the simulated MFT speed at the surface is
always the same as the meridional 
flow-speed used as input to the model, but is different in a two-dimensional
transport model in the meridional ($r,\theta$) plane.
The difference depends
on the value of the magnetic diffusivity 
and on the radial gradient of the latitudinal velocity. 
We have confirmed our results with two different codes in spherical
and Cartesian coordinates.
\end{abstract}


\section{Introduction}

At the solar surface magnetic features are observed in the form 
of active regions. Tracking the motion of 
such a structure individually, one finds in general a poleward migration
which is suggestive of a poleward meridional flow at and just beneath
the Sun's surface.
However, Doppler measurements of the poleward flow speed
at the surface reveal a systematic difference from the speed inferred  
from magnetic feature-tracking (MFT): at low and mid latitudes,
the latter is observed to be lower than the Doppler speed,
but similar to it at high latitudes \citep[see Fig.~10 of][]{ulrich2010}.
Sunspots are usually discarded in such an analysis 
\citep{komm+etal_93,hatha+righ_10} because their motion
may be affected by their strong magnetic fields.
To understand the physical origin of these
differences, \citet{dgu2010} performed a simple test using a
2D (axisymmetric) advective-diffusive flux-transport model.
They showed in simulations  that, due to diffusive transport, the
MFT speed can indeed be different from that of the meridional flow
fed into the model. 
They attributed this difference to the latitudinal
gradient of the  radial component of the magnetic field,
directed towards the equator at the equatorward side of
an active region and towards the pole at its poleward side.
They concluded that magnetic features drift poleward with a net 
speed that is lower than the flow speed at low latitudes and higher
at high latitudes.

In non-axisymmetric two-dimensional ($\theta,\phi$) 
surface-transport models 
\citep[e.g.][]{baumann+etal_04,sheeley2010, wrs2009} 
one could likewise suppose that diffusion is the only agent that
can prevent magnetic features from simply being advected with 
the meridional flow.  However, differences between  
Doppler and MFT speeds have never been discussed  for
those models.

In this paper, we use both 1D and 2D advective-diffusive
flux transport models to clarify to what extent the value  of 
the magnetic diffusivity and its radial gradient influence the 
difference between the
circulation and MFT speeds. Moreover, we will study the role of 
the radial gradient of the latitudinal flow velocity.

\section{Models and methods}
For the sake of simplicity we consider the evolution of azimuthally 
averaged, purely poloidal, i.e., meridional fields. Quite generally, 
studying averaged fields requires the inclusion of an additional 
mean electromotive force $\meanEMF$ in the induction equation.
Its major constituents are often described by the $\alpha$ effect, 
turbulent pumping and turbulent diffusivity $\etat$ 
\citep[see, e.g.,][]{moffatt_78}.  
However, only the latter will be taken into account here and assumed 
to be isotropic, yet possibly  depending on depth.
We do not claim that all other mean-field effects, 
in particular turbulent pumping, are negligible, but
prefer to clarify the origin of the speed deviations in question
by considering  the effects in isolation.
Thus, we focus here on the competition between diffusion and advection.

Our model is kinematic and we consider axisymmetric solutions of the 
induction equation in spherical coordinates $(r,\theta,\phi)$
\EQ
  \parder{\BB}{t}=\nab\times(\UU\times\BB-\etaT\nab\times\BB), 
\quad \nab\cdot\BB=0, \label{induc}
\EN
with the total (molecular plus turbulent) magnetic diffusivity 
$\etaT = \eta + \etat$ and the {\em prescribed} velocity 
$\UU$, i.e., we disregard the back-reaction of the magnetic field onto
$\UU$.  The computational domain spans over a spherical half-shell 
$R_b \le r \le \rsun$,  $0\le\theta\le\pi/2$ (i.e., from the pole to the
equator), where $\rsun$ is the solar radius and the base of the 
convection zone is  at $R_b=0.7\rsun$. The total diffusivity $\etaT$
is, unless specified otherwise, constant across the domain and 
considered a free parameter of the model.

For the 1D version of the model, we solve the radial part of 
\Eq{induc} for $B_r$ at $r=\rsun$, as done
in several surface-transport models
\citep[see, e.g.,][]{devore+etal_84,baumann+etal_04}:
\begin{equation}
\frac{\partial B_r}{\partial t} = -\frac{1}{\rsun \sin\theta}
\frac{\partial}{\partial \theta}\left[\sin\theta\left(U_{\theta}B_r
-\frac{\etaT}{\rsun}\frac{\partial B_r}{\partial \theta}\right)\!\right].
\label{eq:1d}
\end{equation}
Note that this equation is subject to
the simplifying assumption $B_\theta \ll B_r$ at the surface  
\citep[see the Appendix of][]{devore+etal_84},
the consequences of which will be assessed later when discussing 
our results. We solve \Eq{eq:1d} by using a second order finite 
difference scheme with 512 grid points\footnote{Same results are 
obtained for resolutions from 128 to 1024 grid points.}.
The time integration is performed with an implicit 
(Crank-Nicholson) method. 

For the 2D version we solve instead of \Eq{induc} the 
corresponding equation for the $\phi$ component of the vector
potential ${\bm A}=A \ithat{\ee}_{\phi}$, where $\BB=\nab\times\AAA$, and
\begin{equation}
\frac{\partial A}{\partial t} = -\frac{1}{s}({\UU}\cdot \nab)(sA)\,
+\,\etaT\!\left(\nabla^2 -\frac{1}{s^{2}} \right)A, \;\: s=r \sin\theta,
\label{eq:2d}
\end{equation}
again utilizing finite differences (Lax-Wendroff scheme for first 
and centered second order scheme for second derivatives).
For all simulations we use 400$^2$ grid points. A convergence 
analysis revealed that for the global evolution of the magnetic field
a resolution of 128$^2$ grid points is already sufficient.
However, a smoother profile of the estimated tracer velocity 
is obtained with the higher resolution. Time integration is done 
with the ADI method  \citep[for details see][]{gue04}. 
\begin{figure}
\centering
\includegraphics[width=.99\columnwidth]{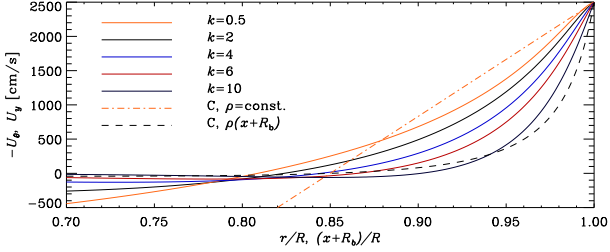}
\caption{Solid lines: Radial profile of 
$U_\theta(r,57^\circ)$ for different 
$k$ as indicated in the legend, 
$n=0.8$, cf. \Eq{eq:F}.
Broken lines: Cartesian velocity profile, $U_y(x,L_y/2)$,
cf. \Eq{eq:Cart}.
}
\label{fig:uthrad}
\end{figure}

In modeling the meridional velocity $\UU$ we start with the 
corresponding mass flow  $\rho\UU$ which is assumed steady and 
has thus to obey $\nab\cdot(\rho\UU)=0$ because of mass conservation.
Hence it can be derived from a stream function $\psi$ by  
$\rho\UU = \nab\times(\psi \ithat{\ee}_{\phi})$.
Following \cite{dgu2010}, we assume an adiabatic density profile
\EQ
\rho(r)=\rho_0(\rsun/r-0.97)^{1.5},   \label{eq:density}
\EN
where $\rho_0$ 
is specified such that 
$\rho(R) \approx {5\cdot10^{-3}}$ g/cm$^3$.
For the stream function we choose the ansatz
$ \psi = - \psi_0 F(r) \partial_\theta G(\theta)$ with
\begin{align}
F(r) &= \left(\frac{\rsun}{r}-0.97 \right)^{n}\!\left[1-\left(\frac{r}{\rsun}\right)^k\right]\!\left[\!\left(\!\frac{R_b}{R}\right)^k\!\!-\left(\frac{r}{R}\right)^k \right]\!,   \label{eq:F}\\  
G(\theta) &= P_2(\cos\theta) + m P_4(\cos\theta),  \label{eq:G} 
\end{align}
where $P_l$ is the Legendre polynomial of order $l$.
$F(r)$ guarantees vanishing $U_r$ at the boundaries $r=R_b,\rsun$ 
assumed impenetrable.
The first factor in $F(r)$, resembling the density profile,
is necessary to avoid local extrema  of $U_{\theta}$ within
the domain which can be achieved by $n=0.8\,\ldots\,1.2$, 
depending on the value of $k$. Apart from that, the exponents
$n$ and $k$ are free 
parameters of the model determining the value of the radial 
derivative of $U_{\theta}$. For fixed $n$,
lower (higher) values of $k$ result in a flatter (steeper)
radial profile $U_{\theta}(r)$ (see Fig.~\ref{fig:uthrad}).
Tuning the exponent $n$ allows adjusting the $U_\theta$ gradient
just at the surface without changing it very strongly deeper down.
So $n$ can be employed for ensuring the stress-free 
boundary condition
$(\partial_r U_{\theta})(R,\theta) = U_{\theta}(R,\theta)/R$,
usually imposed in, e.g., mean-field hydrodynamic models of stellar 
rotation and in direct numerical simulations of convection.
Here, it can be expressed  in the form $h_r(R)=1$,
where  $h_r(r)$ is the normalized radial gradient of
$U_\theta$,  $h_r(r)=r(d\ln F/dr)$.
For simplicity, we ignored this condition in most of our 
calculations  and fixed $n=0.8$. However, we have checked the influence 
of having the stress-free condition obeyed on our results
in a number of cases with different values of $n$. 

Further, the choice of $m=-0.2$ results in a latitudinal surface 
profile $U_{\theta}(\rsun,\theta)$ which resembles 
Doppler velocity observations \citep[see, e.g.,][]{ulrich2010},
in particular the position of the
surface maximum of 
$U_{\theta}(\rsun,\theta)$ is fairly well reproduced. 
$\psi_0$ is adjusted such that 
this maximum is $U_0=2500\,{\rm cm/s}$.

\begin{figure}
\centering
\includegraphics[width=.6\columnwidth]{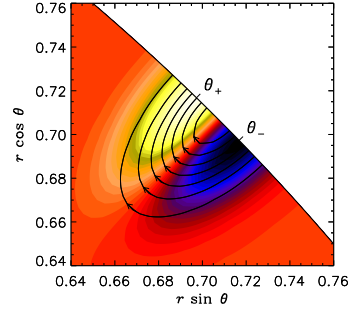} 
\caption{Bipolar region according to \Eq{eq:br} for $\theta_{\rm i}=45^{\circ}$.
$\theta_\pm$ is the position of the centre of the polewards (positive) and 
equatorwards (negative) spot, respectively.
Color coding: $B_r$.
}
\label{fig:br}
\end{figure}

The initial magnetic field of a bipolar region is modelled 
as a flux loop in a meridional plane corresponding to two rings of 
concentrated magnetic flux on the surface of the
sphere. We describe it  by the vector potential
\begin{equation}
\AAA=A_0\exp\left[-\left(\frac{\theta-\thetaI}{w_{\theta}}\right)^2
\right]\exp\left[
-\left(\frac{r-\rsun}{w_r}\right)^2\right]\ithat{\ee}_\phi ,
\label{eq:br}
\end{equation}
where $\thetaI$ is the initial latitudinal location of 
the center of the bipolar region. 
The initial separation between the positions
at which $B_r$ assumes its extrema at the surface, that is, the 
``spot separation" is  $\ga\!\!\!\sqrt{2}\, w_{\theta}$,  
whereas the depth to which the loop extends is controlled by $w_r$.
We assume $w_{\theta}=0.02$ ($2.3^{\circ}$)
and $w_r=0.04\rsun$ throughout this paper. 
For the corresponding field geometry see Fig.~\ref{fig:br}.

Magnetic boundary conditions are chosen to be consistent with 
a perfect conductor in $r\leq R_b$
and  to ensure continuity of $\BB$ with an external 
potential field at $r=\rsun$. 
Comparisons with the simpler normal field condition $\rr\times\BB=\bm{0}$
instead of the potential field condition showed no noticeable 
difference in the results.
At the equator $\BB$ is assumed to be antisymmetric.

Both models were run over the model time interval $T$, being typically 
two weeks, for 20 equidistant initial latitudes $\thetaI$
of the bipolar region between $5^\circ$  and $85^\circ$.  
For measuring the latitudinal surface drift velocity of the flux loop,
averaged over $T$, two different methods were employed. In the first one 
the position  $\theta_{0}$, at which the normal magnetic field $B_r$
vanishes, was followed. 
In the second, we trace the positions of the 
local maximum and minimum of $B_r$ within the loop, 
$\theta_+$  and $\theta_-$, respectively.
The averaged latitudinal velocity  was then defined as  
$\overline{V}_0=\rsun(\theta_{0}(T)-\theta_{0}(0))/T$ in the first case
and as the average of the two values 
$\overline{V}_\pm=\rsun\big(\theta_\pm(T)-\theta_\pm(0)\big)/T$,
that is, 
\EQ
\overline{V} = (\overline{V}_++\overline{V}_-)/2  \label{eq:V} 
\EN
 in the second.
We assign $\overline{V}_0$ to the  average colatitude
$\big(\theta_{0}(T) + \theta_{0}(0)\big )/2$,
but $\overline{V}$ to the average 
$\big(\theta_+(T) + \theta_-(T) + \theta_+(0) + \theta_-(0)\big)/4$.
As the profiles $\mean{V}(\theta)$ and $\mean{V}_0(\theta)$, obtained 
directly in this way, turned out to be rather wiggly we fitted them 
just to $U_\theta(\rsun,\theta)$, that is,
to the function $G$ in \Eq{eq:G}, but with an adjustable amplitude
as fit parameter.
For the highest diffusivity used and for starting 
latitudes $\thetaI$ closest to the equator, the influence of 
this reflecting boundary
becomes noticeable. This influence leads to an unrealistically 
low velocity of the equatorward (negative) spot
which, in turn, corrupts  $\mean{V}$. We use instead $\mean{V}_0$ there.

\section{Results}
\subsection{One-dimensional model}

\begin{figure}
\centering
\includegraphics[width=0.98\columnwidth]{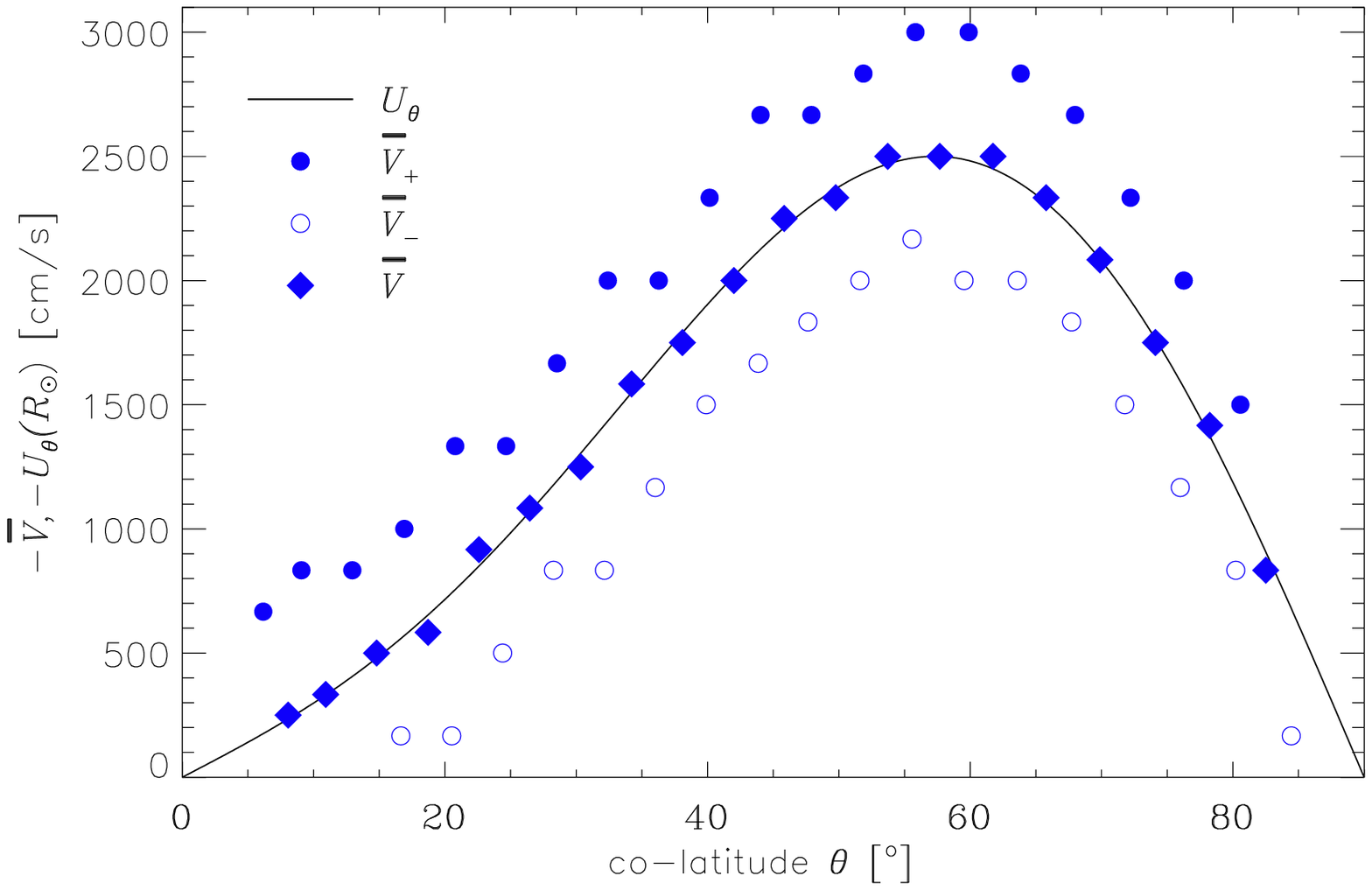} 
\caption{Tracking velocities for the 1D model \eqref{eq:1d} with 
$\etaT=10^{12}\diff$. Solid line: flow velocity
$U_\theta(\rsun)$, filled and open circles: $\overline{V}_+$ 
and $\overline{V}_-\,$, respectively;
diamonds: average $\overline{V}=(\overline{V}_+ + \overline{V}_-)/2$.}
\label{fig:1dvel}
\end{figure}

In simulations with high diffusivity, $\etaT=10^{12}\diff$,
we find that the speed of the poleward
spot, $\overline{V}_+$, is larger than the fluid velocity,
whereas that of the equatorward spot, $\overline{V}_-$,
is smaller; see Fig.~\ref{fig:1dvel}.
However, the average velocity $\overline{V} $ matches  the fluid 
velocity $U_\theta$ fairly well.
The velocity of the center of the bipolar region,  $\overline{V}_0$, also 
coincides with it.
For even higher values of $\etaT$,  $\overline{V}_+$ and $\overline{V}_-$ 
deviate  stronger from 
$U_\theta$, but the average continues to agree with it. 
For evolution times shorter than two weeks 
(e.g., one week or less) the curve for $\overline{V}$ is more wiggly.
However, it always follows the flow. 
These results agree with those of the 2D ($\theta,\phi$) model 
of \cite{wrs2009},  where the poleward spots of 
the bipolar regions move faster than the fluid
for a similar value of $\etaT$ (see their Fig.15).  

\subsection{Two-dimensional model}
\label{sect:2D}
\begin{figure}
\centering
\includegraphics[width=1.\columnwidth]{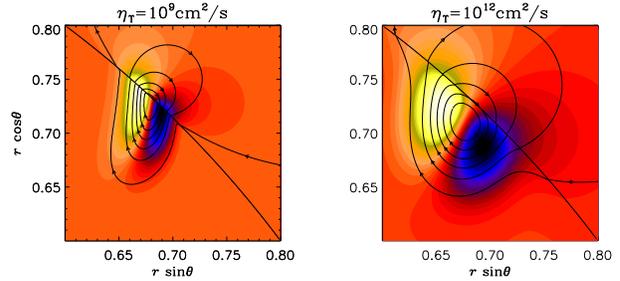} 
\caption{
Magnetic field (solid lines) after two weeks of evolution 
of a bipolar region initially at $\theta_{\rm i}=45^{\circ}$, 
see Fig.~\ref{fig:br}. Color coding: $B_r$. Left: $\etaT=10^9\diff$;
right: $\etaT=10^{12}\diff$ (right).}
\label{fig:br1}
\end{figure}

Next, we study the evolution of a two-dimensional bipolar region
by solving \Eq{eq:2d} using \Eq{eq:br} as initial condition.
As a representative case, Fig.~\ref{fig:br1} shows the evolution 
for a bipolar region initially located at
$\thetaI=45^{\circ}$, using $k=6$ with either
$\etaT=10^9\diff$ (left) or $10^{12}\diff$ (right).

\begin{figure}
\centering
\hspace*{-2.5mm}\includegraphics[width=1.05\columnwidth]{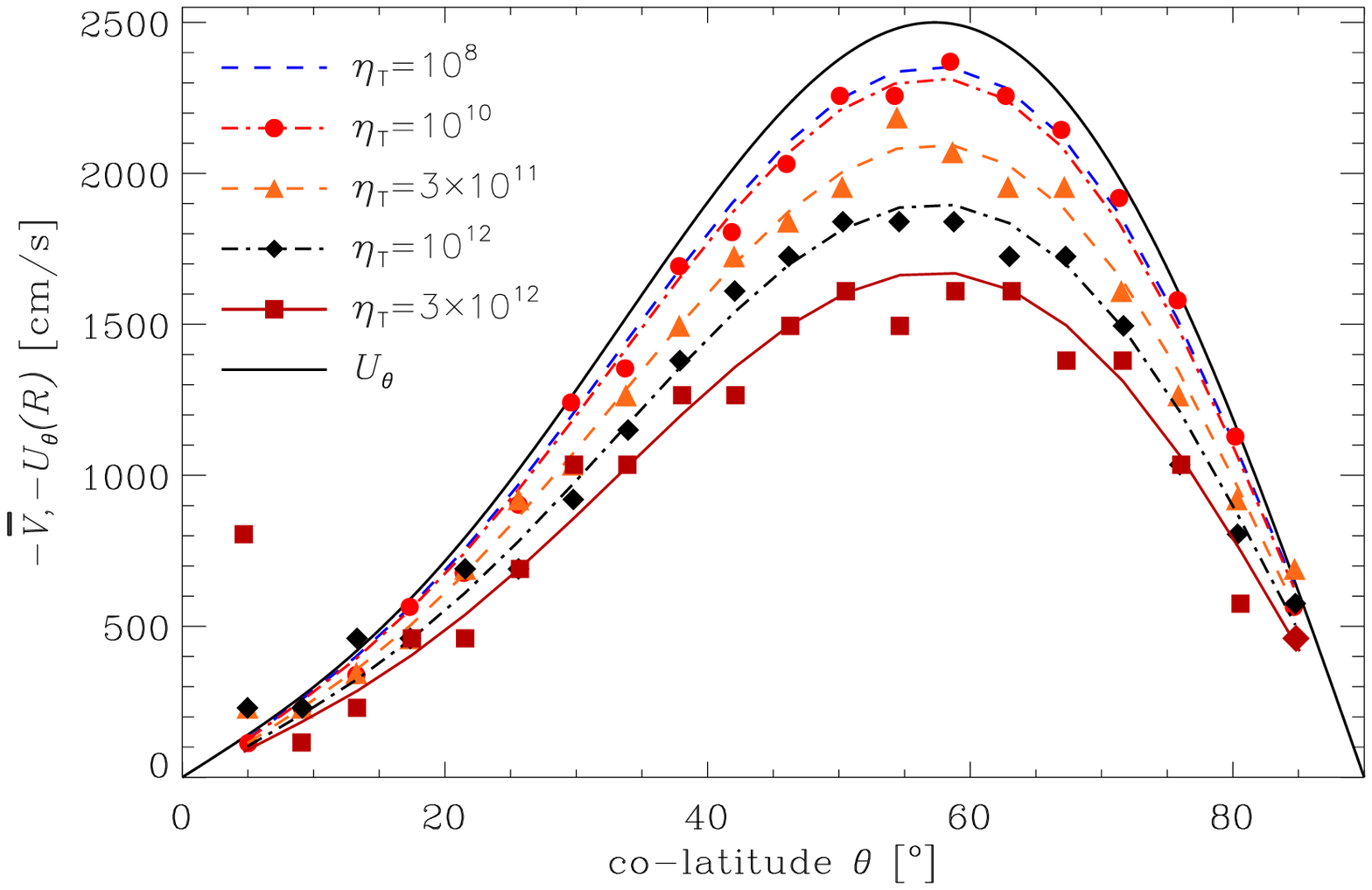} \\
\caption{
Tracking velocity $\mean{V}(\theta)$  for  
$10^8\diff \!\le\!\etaT\! \le \! \!3\times10^{12}\diff$, $k=6$. 
Solid/black: flow velocity $U_\theta(\rsun,\theta)$.
Symbols: $\mean{V}$ according to \Eq{eq:V},
omitted for $\etaT=10^8\diff$ as mostly coinciding with those for 
$\etaT=10^{10}\diff$; symbol for 
$\etaT=3\!\times\!10^{12}\diff$, $\theta=85^\circ$ shows $\mean{V}_0$.
Lines: data fitted to \Eq{eq:G} with amplitude as fit parameter, 
$m=-0.2$ fixed. Colors/symbols/line styles correspond to different values 
of $\etaT$ according to legend. 
\label{fig:vel1w}}
\end{figure}

\begin{figure}
\centering
\hspace*{-2.5mm}\includegraphics[width=1.05\columnwidth]{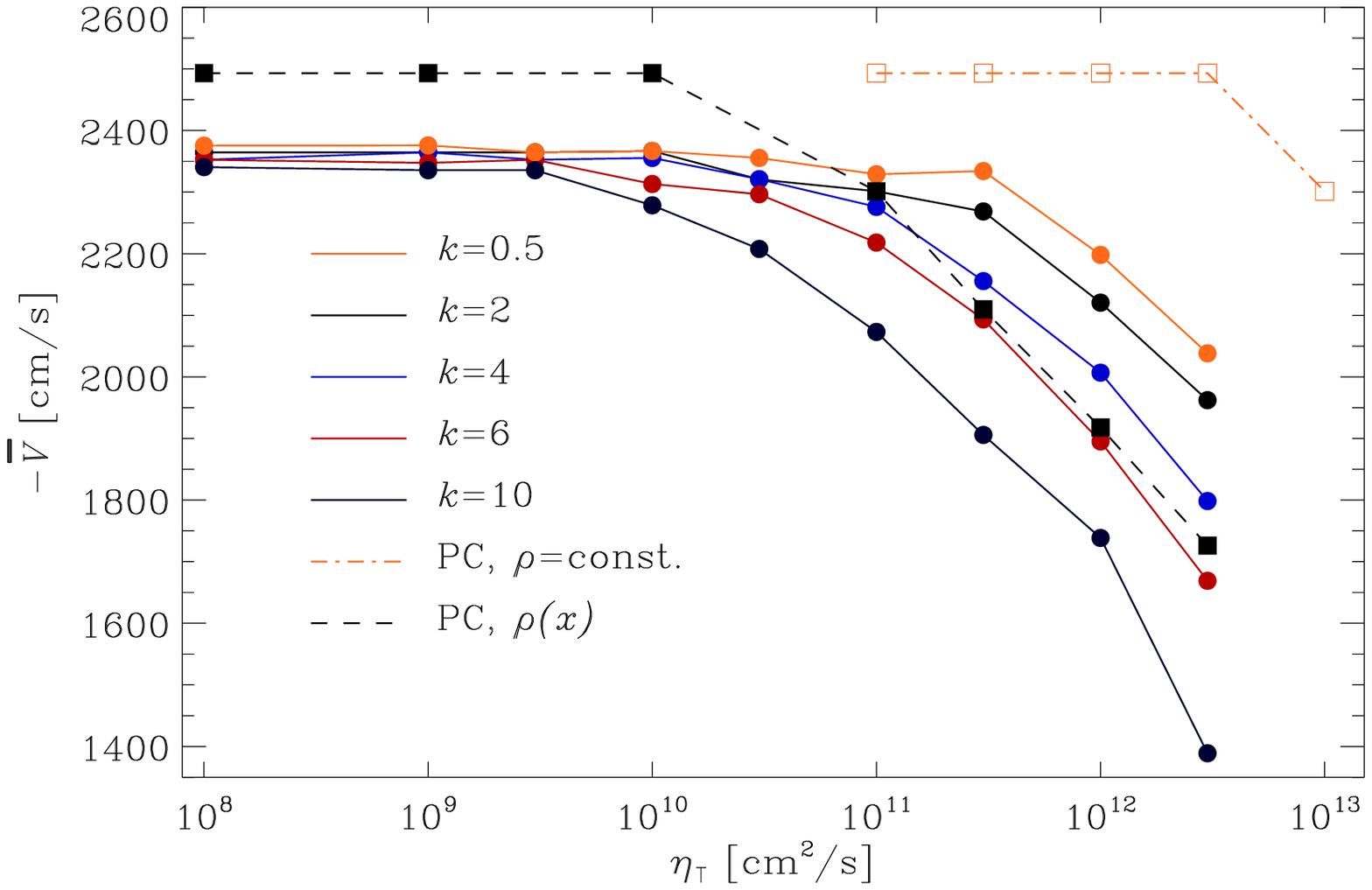} 
\hspace*{-2.5mm}\includegraphics[width=1.05\columnwidth]{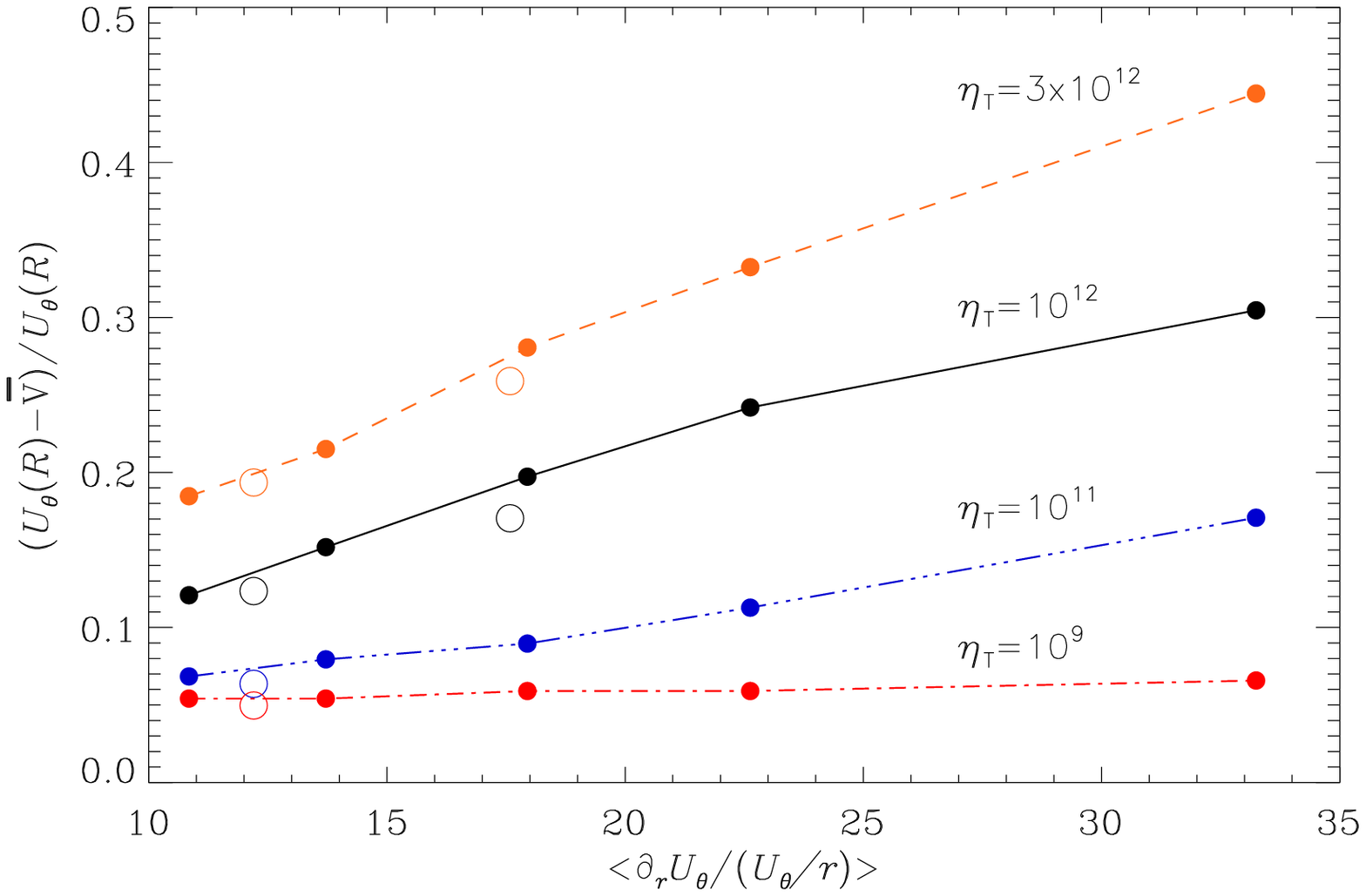} 
\caption{
Upper panel: Maximum tracking velocity $\mean{V}$ vs. $\etaT$ for
different values of $k$ in  \Eq{eq:G}.
Solid lines/filled circles
correspond to the spherical model (values taken from fit curves).
Broken lines/open squares correspond to the Cartesian
model (C); dashed/filled squares: density according to 
\Eq{eq:density}, dash-dot/open squares: constant density.
Lower panel, lines/closed circles: fractional speed difference,
$(U_\theta(R)-\mean{V})/U_\theta(R)$, vs. the normalized radial 
gradient of $U_{\theta}$, $h_r(r)$, averaged over
$r/R=0.97\ldots1$.
Open circles: stress-free boundary 
condition ensured in the profile \eqref{eq:F} by $(n,k)=(0.954,4)$  for $\bra{h_r}\approx 12$
and $(1.194,10)$ for $\bra{h_r}\approx 17.5$.
\label{fig:veleta1w}}
\end{figure}

In contrast to the 1D model, where the 
profile $\mean{V}(\theta)$ turns out to be independent 
of $\etaT$, we find here a significant dependence.
For small values of $\etaT$, the
``frozen-in'' condition is fulfilled and thus the magnetic field 
lines appear indeed attached to the plasma flow.
(The systematic offset between $\mean{V}$ and 
$U_\theta$ for $\etaT\rightarrow 0$,
visible in Figs.~\ref{fig:vel1w} and \ref{fig:veleta1w},
is mainly due to the discretization errors.)
For larger $\etaT$ ($\ga 3\times10^{11}\diff$), 
however, the diffusion time scale becomes similar to or even 
smaller than the  advection time scale,
and then there is an increasing departure from the ``frozen-in'' state.
This becomes clear in Fig.~\ref{fig:vel1w}, where
$\overline{V}(\theta)$  is shown for 
$10^8 \le \etaT\le 3\times10^{12}\diff$. 
In general,  the deviation from $U_\theta$ increases everywhere with 
growing $\etaT$, while for each $\etaT$  it adopts its largest value 
at intermediate latitudes of $\approx 57^\circ$
where $U_\theta$ peaks.
Note that the simple fit based only on the amplitude,
using the function $G$ from \Eq{eq:G}, works remarkably well.


The top panel of Fig.~\ref{fig:veleta1w} visualizes the dependence 
of $\mean{V}$ on the radial variation of $U_{\theta}$, i. e., on the 
index $k$ in \Eq{eq:G}. We present $\mean{V}$ at the latitude
where $U_\theta$ peaks ($\theta\approx57^{\circ}$) as
a function of $\etaT$ for $k$ varying from $0.5$ to $10$.
To minimize the effect of numerical noise we have taken
$\mean{V}$ from fit curves.
For $k=0.5$ (yellow line), 
$\mean{V}$ does not depend on $\etaT$ up to $10^{11}\diff$.
Beyond this value, $\mean{V}$ starts to decrease. 
For increasing $k$ the curves depart from the ``frozen-in'' 
domain at decreasing values of $\etaT$ being as small as 
$\approx3\times10^9\diff$ for $k=10$.
The bottom panel of the same figure shows the fractional 
velocity difference,
$\big(U_\theta(R)-\mean{V}\big)/U_\theta(R)$,
as a function of the normalized radial gradient of $U_{\theta}$,
$h_r(r)$, averaged over the  interval $r=0.97R\,\ldots\,R$,
where the major part of the magnetic flux is residing.
Note that for the highest diffusivity, $3\times10^{12}\diff$,
$\mean{V}$ is reduced by $\approx\!45$\% at $\bra{h_r}\approx33$.
Employing our results for interpreting the data given
in Fig.~10 of \citet{ulrich2010}, we find that
their speed reductions of about 30\% do
occur in our model, either for 
$\etaT=3\times10^{12}\diff$ and $\bra{h_r}\approx20$
or for $\etaT=10^{12}\diff$ and $\bra{h_r}\approx33$. 
In the bottom panel of Fig.~\ref{fig:veleta1w}  some results are
shown with the profile \eqref{eq:F} adjusted to the stress-free boundary 
condition by fine-tuning of $n$. Obviously, there is only a slight 
reduction of the velocity difference in comparison with the unadjusted 
profile.

\subsection{Dependency on $\etaT(r)$}

Having examined the influence of the radial profile of 
$U_\theta$ on $\mean{V}$,
one must ask whether also
the radial profile of $\etaT$ has an effect.
From an observational point of view this profile
is unknown.
Hence, the profiles so far considered in dynamo models
are to some extent arbitrary. For instance, \cite{dikgil01} and \cite{gue07a} 
have used a step function,
with amplitudes of  $10^{10}\diff$ in the bulk of the convection 
zone and a value of $\approx10^{12}\diff$ for supergranular diffusion 
within the so-called near-surface shear layer.  On the other hand,
\cite{pipin+etal_11} considered
a mixing length theory (MLT) estimation of $\etat$.  

\begin{figure}
\centering
\includegraphics[width=0.99\columnwidth]{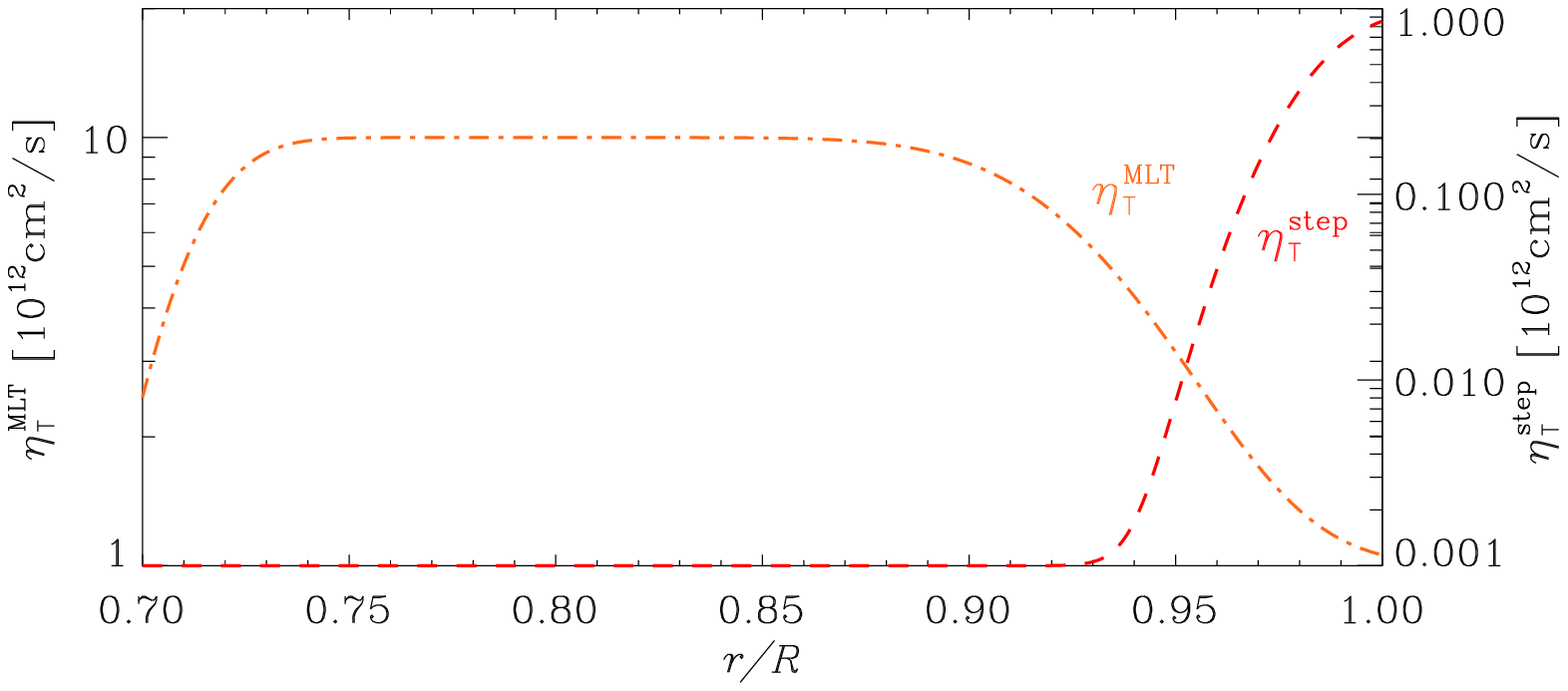}
\includegraphics[width=0.99\columnwidth]{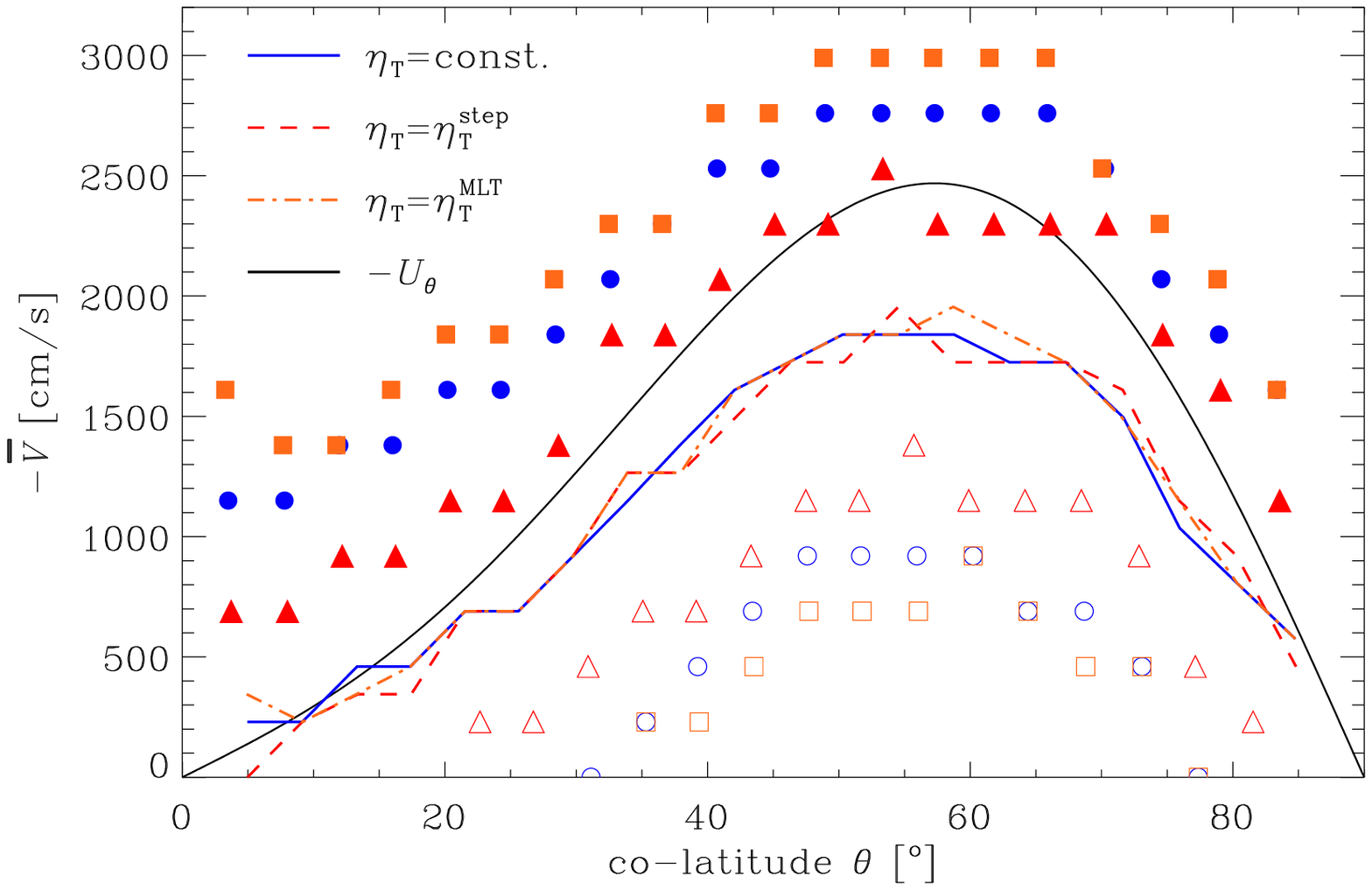}
\caption{Top:
Profiles of $\etaT$ used. Bottom:
Corresponding tracking velocities $\mean{V}$  for
$\etaT(\rsun) = 10^{12}\diff$, $k=6$. 
Solid/black: flow velocity $U_\theta(\rsun)$.
Filled (open) symbols: $\mean{V}_+(\mean{V}_-)$.
}
\label{fig:diffeta}
\end{figure}

Here we consider both a step and an MLT profile defined 
by the following  expressions (see the profiles in the top panel 
of Fig.~\ref{fig:diffeta}):
\begin{equation}
\etaT^{\rm step}=\eta_{\rm cz}+\frac{\eta_{\rm s}-\eta_{\rm cz}}{2}\left[1+{\rm erf}
  \left(\frac{r-0.96\rsun}{0.02\rsun} \right) \right],
\end{equation}
with $\eta_{\rm s}=10^{12}\diff$ and $\eta_{\rm cz}=10^{-2}\eta_s$, and
\begin{alignat}{2}
\etaT^{\rm MLT}=\eta_{\rm rz}&+\frac{\eta_{\rm cz}-\eta_{\rm rz}}{2}&&\left[1+{\rm erf}
  \left(\frac{r-0.71\rsun}{0.02\rsun} \right) \right]  \\\nonumber
&+\frac{\eta_{\rm s}-\eta_{\rm cz}}{2}&&\left[1-{\rm erf}
  \left(\frac{r-0.93\rsun}{0.04\rsun} \right) \right] ,
\end{alignat}
where  $\eta_{\rm rz}=10^{8}\diff$,  $\eta_{\rm s}=10^{13}\diff$, and 
$\eta_{\rm cz}=10^{-1}\eta_{\rm s}$.

We have performed numerical experiments with $k=6$
for $U_\theta$ and a fixed surface value $\etaT(\rsun)=10^{12}\diff$. 
The results, displayed in the bottom panel of 
Fig.~\ref{fig:diffeta}, do not show marked differences
between the three diffusivity profiles considered, 
and the MFT velocity profile is roughly the same.
However, the separation
between poleward and equatorward spots is smaller
for the model with the step profile.
The model with constant $\etaT$ exhibits an intermediate
difference whereas in the model with the MLT
profile the dispersion increases.

\subsection{Comparison with Cartesian geometry}
To assess the influence of the curvature in our spherical model 
we have repeated some of the simulations in a 2D ($L_x \times L_y$) 
Cartesian box with aspect ratio $(R-R_b) : R\,\pi/2$
and a simplified circulation velocity,
\EQ
\hspace*{-1mm}\UU = \frac{1}{\rho}\nab\times(\psi \ithat{\ee}_z), \;\:
\psi=\psi_0 \frac{x}{L_x}\!\!\left(\frac{x}{L_x}\!  -1 \right)
\frac{y}{L_y}\!\!\left(\frac{y}{L_y} -1 \right)\!,  \label{eq:Cart}
\EN
with
($x$, $y$) corresponding to ($r$, $\theta$), respectively,  
$\rho$ set either constant or to $\rho(x+R_b)$ 
from \Eq{eq:density},
and $\psi_0$ again adjusted to yield $2500\,{\rm cm/s}$ for the maximum 
surface velocity. Instead of the vacuum boundary condition at $r=\rsun$, 
the simpler normal field condition $\ithat{\ee}_x\times\BB=\bm{0}$ was 
imposed there. These simulations were performed with the 
{\sc Pencil Code}\footnote{\texttt{http://pencil-code.googlecode.com/}},
which uses sixth-order explicit finite differences in space and third
order accurate time stepping method. For these runs we use 512$^2$
grid points.

The results, here with constant $\etaT$, for both choices of $\rho$,
are represented by broken lines in Fig.~\ref{fig:veleta1w}.
The corresponding profiles $U_y(x,L_y/2)$ are shown in 
Fig~\ref{fig:uthrad} with the same line styles. Clearly, these 
findings agree with those of the spherical model showing that the 
value of $\etaT$, at which the frozen-in condition starts to fail, 
depends crucially on the radial
($x$) gradient of $U_y$ and varies here 
by about  two orders of magnitude.
\section{Discussion and Conclusions}
Through 1D and 2D advective-diffusive models we have investigated
the differences between the surface meridional 
flow speed obtained from Doppler measurements and that inferred 
from magnetic feature-tracking. In the one-dimensional simulations,
the average velocity of the magnetic tracers always coincides with that
of the flow, independently of the value of $\etaT$.
In 2D models, on the other hand, flow and 
feature-tracking velocities may diverge at higher diffusivities,
for which the ``frozen-in'' condition does no longer hold.
Further, the difference between these velocities depends on the
radial gradient of the latitudinal velocity:
the steeper $U_{\theta}$, the larger the difference.
Using a different code we have verified
that these results apply also
in Cartesian geometry. To understand this dependence we 
refer to the induction equation (for simplicity in Cartesian coordinates),
taken at the surface  $x=L_x$ (or $r=R$) where $U_x=\partial_y U_x=0$, so
\begin{alignat}{2}
  \label{eq:bx}
  \frac{\partial B_x}{\partial t}&=-&&\frac{\partial}{\partial y}
  (U_yB_x) + \etaT\left(\frac{\partial^2 B_x}{\partial x^2} + 
  \frac{\partial^2 B_x}{\partial y^2}  \right),\\
  \label{eq:by}
\frac{\partial B_y}{\partial t}&=&&\frac{\partial}{\partial x}
(U_yB_x-U_xB_y)+\etaT\left(\frac{\partial^2 B_y}{\partial x^2} + 
  \frac{\partial^2 B_y}{\partial y^2}  \right).
\end{alignat}
Note that $B_x$ is apparently decoupled from $B_y$, but at 
the price of being coupled to its second vertical derivative.
In the case
of small $\etaT$ (i.e., diffusion time larger than advection time),
the evolution of $B_x$ is governed by the first, advective, 
term in \Eq{eq:bx}. In this case the magnetic field lines
follow the fluid velocity locally
(see curved magnetic field lines in the 
left hand panel of Fig. \ref{fig:br1}). 
In the case of larger $\etaT$ 
(diffusion time similar to or shorter than
advection time), the diffusion term in \Eq{eq:bx}
plays a significant role in the field evolution.
The dependence on $\partial_x^2 B_x$ can be eliminated 
by the solenoidal condition
$\partial_x B_x+\partial_y B_y =0$, by which
the coupling between the two equations becomes obvious.
Because $B_y$ depends explicitly on $\partial_x U_y$,
the surface speed of $B_x$ will clearly be
modified by the fluid motion deeper down in the sub-surface layers.
When ignoring $B_y$ from the beginning, however, as done in
\cite{devore+etal_84}, and many surface transport models afterwards,
this influence will be lost.
In spherical coordinates, the induction equation for $B_r$ 
exhibits an analogous dependence on $B_\theta$,
hence the same argument is valid.

Based upon our results for the difference between
flow and feature-tracking speeds, one might think of inferring
the thickness of the layer where the flow is poleward.
This value, however, would depend on the surface diffusivity
and on $\partial_r U_{\theta}$, both of
which are poorly known.
\cite{hathaway_11} has inferred an extreme flow pattern with a very
shallow poleward flow ($\approx35\Mm$ deep) which nevertheless
can be brought into agreement with our findings, requiring
a large radial  gradient of $U_\theta$,
that is, $\bra{h_r(r)}\ga 20$, cf. Fig.~\ref{fig:veleta1w}.
On the other hand, surface
flux-transport models in two dimensions  ($\theta,\phi$)
which disregard the radial derivatives in $B_r$,
are probably overestimating the importance of advection
in their results.

\section*{Acknowledgments}
This work started during the NORDITA program on 
predictability and data assimilation and is supported
by the European Research Council under the AstroDyn
research project 227952. MD thanks the support by NASA's 
Living With a Star program through the grant NNX08AQ34G.

\def\apj{ApJ}
\def\aap{A\&A}
\def\mnras{MNRAS}
\def\solphys{Sol. Phys.}
\bibliographystyle{mn2e}
\bibliography{bib}

\begin{thebibliography}{}

\bibitem[\protect\citeauthoryear{{Baumann}, {Schmitt}, {Sch{\"u}ssler} \&
  {Solanki}}{{Baumann} et~al.}{2004}]{baumann+etal_04}
{Baumann} I.,  {Schmitt} D.,  {Sch{\"u}ssler} M.,    {Solanki} S.~K.,  2004,
  A\&A, 426, 1075

\bibitem[\protect\citeauthoryear{{DeVore}, {Sheeley} Jr. \& {Boris}}{{DeVore}
  et~al.}{1984}]{devore+etal_84}
{DeVore} C.~R.,  {Sheeley} Jr. N.~R.,    {Boris} J.~P.,  1984, Sol. Physics,
  92, 1

\bibitem[\protect\citeauthoryear{{Dikpati} \& {Gilman}}{{Dikpati} \&
  {Gilman}}{2001}]{dikgil01}
{Dikpati} M.,  {Gilman} P.~A.,  2001, ApJ, 559, 428

\bibitem[\protect\citeauthoryear{{Dikpati}, {Gilman} \& {Ulrich}}{{Dikpati}
  et~al.}{2010}]{dgu2010}
{Dikpati} M.,  {Gilman} P.~A.,    {Ulrich} R.~K.,  2010, ApJ, 722, 774

\bibitem[\protect\citeauthoryear{{Guerrero} \& {de Gouveia Dal
  Pino}}{{Guerrero} \& {de Gouveia Dal Pino}}{2007}]{gue07a}
{Guerrero} G.,  {de Gouveia Dal Pino} E.~M.,  2007, A\&A, 464, 341

\bibitem[\protect\citeauthoryear{{Guerrero} \& {Mu{\~n}oz}}{{Guerrero} \&
  {Mu{\~n}oz}}{2004}]{gue04}
{Guerrero} G.~A.,  {Mu{\~n}oz} J.~D.,  2004, MNRAS, 350, 317

\bibitem[\protect\citeauthoryear{{Hathaway}}{{Hathaway}}{2011}]{hathaway_11}
{Hathaway} D.~H.,  2011, arXiv:1103.1561

\bibitem[\protect\citeauthoryear{{Hathaway} \& {Rightmire}}{{Hathaway} \&
  {Rightmire}}{2010}]{hatha+righ_10}
{Hathaway} D.~H.,  {Rightmire} L.,  2010, Science, 327, 1350

\bibitem[\protect\citeauthoryear{{Komm}, {Howard} \& {Harvey}}{{Komm}
  et~al.}{1993}]{komm+etal_93}
{Komm} R.~W.,  {Howard} R.~F.,    {Harvey} J.~W.,  1993, \solphys, 147, 207

\bibitem[\protect\citeauthoryear{{Moffatt}}{{Moffatt}}{1978}]{moffatt_78}
{Moffatt} H.~K.,  1978, {Magnetic field generation in electrically conducting
  fluids}.
Cambridge Univ.\ Press, Cambridge

\bibitem[\protect\citeauthoryear{{Pipin}, {Kuzanyan}, {Zhang} \&
  {Kosovichev}}{{Pipin} et~al.}{2011}]{pipin+etal_11}
{Pipin} V.~V.,  {Kuzanyan} K.~M.,  {Zhang} H.,    {Kosovichev} A.~G.,  2011,
  arXiv:1105.4285

\bibitem[\protect\citeauthoryear{{Sheeley} Jr.}{{Sheeley}}{2010}]{sheeley2010}
{Sheeley} Jr. N.~R.,  2010, in {S.~R.~Cranmer, J.~T.~Hoeksema, \& J.~L.~Kohl}
  ed., SOHO-23: Understanding a Peculiar Solar Minimum Vol.~428 of Astronomical
  Society of the Pacific Conference Series, {What's So Peculiar about the Cycle
  23/24 Solar Minimum?}.
p.~3

\bibitem[\protect\citeauthoryear{{Ulrich}}{{Ulrich}}{2010}]{ulrich2010}
{Ulrich} R.~K.,  2010, ApJ, 725, 658

\bibitem[\protect\citeauthoryear{{Wang}, {Robbrecht} \& {Sheeley}}{{Wang}
  et~al.}{2009}]{wrs2009}
{Wang} Y.-M.,  {Robbrecht} E.,    {Sheeley} N.~R.,  2009, ApJ, 707, 1372

\end{thebibliography}

\label{lastpage}

\end{document}